\documentclass[prl,twocolumn,amsmath,amssymb]{revtex4}

\begin{document}

\title{Very Special Relativity}

\date{Jan 26, 2006}

\author{Andrew G. Cohen}
\email{cohen@bu.edu}
\author{Sheldon L. Glashow}
\email{slg@bu.edu}
\affiliation{Physics Department,  Boston University\\
  Boston, MA 02215, USA}

\begin{abstract}
  By Very Special Relativity (VSR) we mean descriptions of nature
  whose space-time symmetries are certain proper subgroups of the
  Poincar\'e group. These subgroups contain space-time translations
  together with at least a 2-parameter subgroup of the Lorentz group
  isomorphic to that generated by $K_{x}+J_{y}$ and $K_{y}-J_{x}$.  We
  find that VSR implies special relativity (SR) in the context of
  local quantum field theory or of $CP$ conservation.  Absent both of
  these added hypotheses, VSR provides a simulacrum of SR for which
  most of the consequences of Lorentz invariance remain wholly or
  essentially intact, and for which many sensitive searches for
  departures from Lorentz invariance must fail.  Several feasible
  experiments are discussed for which Lorentz-violating effects in VSR
  may be detectable.
\end{abstract}

\maketitle


Special relativity (SR) is based on the hypothesis that the laws of
physics share many of the symmetries of Maxwell's equations.  Whereas
the maximal symmetry group of Maxwell's equations is the 15-parameter
conformal group $SU(2,4)$, the existence of particles with mass (and
the known violations of $P$ and $T$) constrains space-time symmetry to
be no greater than the Poincar\'e group (the connected component of
the Lorentz group along with space-time translations). The special
theory of relativity identifies this group as the symmetry of nature.

Although no decisive departure from exact Lorentz invariance has yet
been detected, ever more sensitive searches should be and are being
carried out. A perturbative framework has been developed to
investigate a certain class of departures from Lorentz invariance. For
example, Coleman and Glashow\cite{Coleman:1998ti,Coleman:1997xq}
consider the case of space-time translations along with exact
rotational symmetry in the rest frame of the cosmic background
radiation, but allow small departures from boost invariance in this
frame.  Perturbative departures from Lorentz-invariance are then
readily parametrized in terms of a fixed time-like 4-vector or
`spurion.'  Others\cite{Colladay:1996iz,Colladay:1998fq} consider the
introduction into the Lagrangian of more general spurion-mediated
perturbations (sometimes referred to as `expectation values of Lorentz
tensors following spontaneous Lorentz breaking.').

In this note we pursue a different approach to the possible failure of
Lorentz symmetry. We ask whether the exact symmetry group of nature
may be isomorphic to a proper subgroup of the Poincar\'e group. To
preserve energy and momentum conservation, we consider only those
subgroups that include space-time translations along with a proper
subgroup of the Lorentz group.  Up to isomorphism and possible
discrete elements, there are four such distinct subgroups with
1-parameter, three with 2-parameters, five with 3-parameters, and just
one with 4-parameters.

For reasons soon to be explained, we restrict our attention to those
subgroups of the Lorentz group containing the generators $T_{1} \equiv
K_{x}+J_{y}$ and $T_{2} = K_{y} - J_{x}$, where $\mathbf{J}$ and
$\mathbf{K}$ are the generators of rotations and boosts
respectively. These commuting generators form a group, $T(2)$, which
is isomorphic to the group of translations in the plane. Three larger
subgroups of the Lorentz group are obtained by adjoining one or two
generators to $T(2)$.  Each of them has a natural action on the plane:
the addition of $J_{z}$ yields a group isomorphic to the
three-parameter group of Euclidean motions, $E(2)$; the addition of
$K_{z}$ yields one isomorphic to the three-parameter group of
orientation-preserving similarity transformations, or homotheties,
$HOM(2)$; and lastly, the addition of both $J_{z}$ and $ K_{z}$ yields
one isomorphic to the four-parameter similitude group,
$SIM(2)$\footnote{There is a further one-parameter family of
  three-dimensional groups in which we adjoin the generator $K_{z}+a
  J_{z}$ for any real, non-zero $a$. We do not consider these here.}.

We refer to any scheme whose space-time symmetries consist of
translations along with any one of the Lorentz subgroups described
above as Very Special Relativity (VSR). We shall see that the four VSR
avatars thus defined have quite different character. Nevertheless,
they all share the following remarkable defining property: that the
incorporation of either $P$, $T$ or $CP$ enlarges these subgroups to
the full Lorentz group.  Conjugation by one of these discrete
transformations treats boosts and rotations oppositely, thereby
extending the group to allow boosts and rotations in the $x$--$y$
plane independently. Further commutation leads to the remaining
$z$-boost and $z$-rotation.  The group $T(2)$ is the smallest subgroup
of the Lorentz group with this property, the only others being those
containing $T(2)$, hence our focus.  It follows that Lorentz-violating
effects in VSR are absent for theories conserving any one of these
three discrete symmetries (and perhaps, that Lorentz-violating effects
in VSR are necessarily small because $CP$ violating effects are
small.)

In previous approaches, the breaking of Lorentz symmetry was expressed
in terms of local operators incorporating one or more invariant
tensors, or spurions. In the case of $SO(3)$ symmetry, the spurion
takes the form of a time-like 4-vector, and the lowest dimension
operators involving it affect both particle propagation and the
kinematics of particle decays. The limits on such departures from SR
in this model are exceptionally strong.  For example, the mere
observation of ultra-high energy cosmic rays places an upper bound of
$10^{-23}$ on one dimensionless measure of Lorentz
violation\cite{Coleman:1998ti}, while an analysis of neutrino data
bounds flavor-dependent Lorentz violation in the neutrino sector to
less than $10^{-25}$\cite{Battistoni:2005gy}.

We may attempt to apply a similar spurion strategy to the four VSR
groups described above. The smallest group $T(2)$ admits many possible
invariant tensors. The simplest of these whose little group is no
greater than $T(2)$ is the antisymmetric two-index tensor\footnote{For
  a finite-dimensional representation of the Lorentz group labeled by
  two non-negative half-integers $(n,m)$, such an invariant tensor
  corresponds to the state with weight $\vert{}n,-m\rangle{}$. To
  avoid invariance under the larger group $E(2)$ $n$ must differ from
  $m$.}
\begin{equation*}
  F =
  \begin{pmatrix}
    0& 1& 0& 0 \\
    -1& 0& 0& -1 \\
    0& 0& 0& 0 \\
    0 &1& 0& 0
  \end{pmatrix}.
\end{equation*}
($F$  may be thought of as the field-strength for a zero frequency
electromagnetic wave with linear polarization in the $x$-direction.)

The group $E(2)$ admits the 4-vector $n=(1,0,0,1)$ as an invariant
tensor.  (This is also an invariant tensor for $T(2)$, but one which
preserves rotations about the $z$-axis as well.) The existence of
invariant tensors for the VSR groups $T(2)$ and $E(2)$ allows the
construction of Lorentz-violating local operators that, among other
things, affect the propagation of particles, much as in the $O(3)$
case. However unlike that case, these new operators necessarily
violate $P$, $CP$ and $T$.

The remaining two VSR groups $HOM(2)$ and $SIM(2)$ are entirely
different in this regard.  There are no invariant tensors for these
cases\footnote{For an irreducible representation labeled by two
  half-integers $(n,m)$ the only state annihilated by $T_{1}$ and
  $T_{2}$ is $\vert n,-m\rangle$. But this is an eigenstate of $K_{z}$
  with eigenvalue $n+m$ which vanishes only for $n=m=0$.}.  No local
Lorentz symmetry-breaking operator preserving either of these groups
exists and there is no obvious local, perturbative description of
their departures from SR.  Consequently, spurions cannot access
scenarios in which the symmetry group of nature is  $HOM(2)$ or
$SIM(2)$. 

The situation for these  groups is much like that of $CPT$
in the context of Lorentz-invariant local quantum field theory: all
local operators preserving Lorentz invariance preserve a larger
symmetry (Lorentz plus $CPT$). Here, all local operators preserving
$SIM(2)$ (or $HOM(2)$) also preserve a larger symmetry
(Lorentz). Nevertheless it is easy to construct non-local amplitudes
that violate Lorentz invariance while respecting $SIM(2)$ (or
$HOM(2)$).

One way to do this makes use of the non-invariant null vector $n\equiv
(1,0,0,1)$. This vector is invariant under $T_{1}, T_{2}$
transformations and $z$-axis rotations, but transforms as $n \to
e^{\phi}n$ under boosts in the $z$-direction.  Consequently, ratios of
dot-products of this vector with kinematic variables (such as momenta)
are invariant under $SIM(2)$ or $HOM(2)$ but not under all Lorentz
transformations. For example, the amplitude for the two body decay of
a spinless particle at rest can depend on the 4-momenta of the decay
products, $p_{1} \text{ and } p_{2}$. The ratio $(p_{1}\cdot
n)/(p_{2}\cdot n)$ is then an invariant, and thus the amplitude for
the decay may depend on the direction of the decay products relative
to the VSR-preferred direction (nominally, the $z$ axis).

Because VSR includes space-time translations, particle states may be
labeled by their 4-momenta. For $SIM(2)$ and $HOM(2)$, the only
invariant that can be constructed from the 4-momentum of a massive
particle is the mass itself, just as in SR. Therefore, all positive
energy time-like momenta of fixed length are equivalent under $SIM(2)$
or $HOM(2)$ transformations. (A given time-like momentum may be
transformed to the rest frame by three successive transformations: a
$T_{1}$ transformation with parameter $-p_{x}/(E-p_{z})$; a $T_{2}$
transformation with parameter $-p_{y}/(E-p_{z})$; and a boost in the
$z$-direction with parameter $e^\phi = (E-p_{z})/M$.)  This result
implies that many of the elementary consequences of SR, such as
time-dilation, the law of velocity addition, the existence of a
center-of-mass frame, and a universal and isotropic maximal attainable
velocity hold in these variants of VSR. Indeed, invariance under
$HOM(2)$, rather than (as is often taught) the Lorentz group, is both
necessary and sufficient to ensure that the speed of light is the same
for all observers, and \emph{inter alia}, to explain the null result
to the Michelson-Morley experiment and its more sensitive successors.

Nature seems well described by the Standard Model, a Lorentz-invariant
local quantum field theory in which the existence of three fermion
families is necessary for $CP$ violation. As a result, $CP$ violating
effects are usually small and are nearly absent in all flavor-diagonal
processes. In this context, we note that the failure to detect the
neutron electric dipole moment shows that $\bar{\theta} < 10^{-10}$,
while arguments suggest that $\bar{\theta}$ may be considerably
smaller.

Were $CP$ an exact symmetry, VSR would imply SR.  Consequently,
because $CP$ violating effects in nature are small, we expect VSR
departures from SR to be correspondingly small.  However, such
departures may be a dominant effect in processes for which $CP$
violation is significant. For example, in the decay $K_L\rightarrow
\pi^++\pi^-$, the pions need not be isotropically distributed in the
kaon rest frame. Their directions could be correlated to the
VSR-preferred direction. This effect is likely to be tiny for
the spurion-inaccessible variants of VSR, because $CP$ violation in
kaon decay is predominantly indirect (propagator dominated) and as we
have noted, particle propagation is unaffected for $SIM(2)$ or
$HOM(2)$.

For the spurion-inaccessible variants of VSR, observable departures
from SR might be found in studies of the $CP$-violating decays of
neutral $B$ mesons (such as $B_0\rightarrow J/\Psi+ K_s$), where $CP$
violation is largely direct and significant angular correlations may
be present.  A straightforward search for departures from SR can be
performed without knowing the times of individual events, so long as
the VSR preferred direction $\vec n$ is not coincident with the
Earth's polar axis.  In that case, we anticipate an angular
distribution (in the $B$ rest frame) depending on the angle between a
decay product and the polar axis.

VSR may have radical consequences for neutrino physics. Neutrinos are
now known to have mass. Several mechanisms have been contrived to
remedy the absence of neutrino mass in the Standard Model. All of
these invoke new particles or new interactions. In the `Dirac'
picture, lepton number is conserved with neutrinos acquiring mass via
(anomalously small) Yukawa couplings to sterile $SU(2)$-singlet
neutrinos. In the `Majorana' picture, lepton number is
violated. Neutrino masses result from a seesaw mechanism involving
heavy sterile states, or via dimension-6 operators resulting from
unspecified new interactions.

In VSR, neutrino mass has a natural origin. Lepton-number conserving
neutrino masses, although not Lorentz invariant, are VSR invariant.
There is no guarantee that neutrino masses have a VSR origin,
but if so their sizes may be an indication of the magnitude of
Lorentz-violating effects in other sectors. For example VSR allows for
an (anisotropic!) electric dipole moment for charged leptons. $SU(2)$
invariance may then relate such dipole moments to neutrino masses:
$d_{\text{lepton}} \sim (m_{\nu}/2m_{l})^{2}(e/2m_{l})$.  For the
electron and for $m_{\nu}^{2}\simeq 10^{-4}\text{ eV}^{2}$ this is the
same size as the current experimental
sensitivity\cite{Regan:2002ta}. We leave detailed explication of these
and related matters to a subsequent publication.

A VSR origin of neutrino masses requires no additional states and need
not introduce lepton number violation\footnote{The massless and
  massive unitary representations of $SIM(2)$ and $HOM(2)$ are
  one-dimensional, unlike those of the Poincar\'e group.}. This is a
significant departure from conventional notions. However, because all
observable neutrino phenomena involve ultra-relativistic neutrinos
($\gamma >> 1$), neutrino phenomenology is virtually identical to that
of the usual scenarios: the neutrino helicity will differ
significantly (but unobservably) from $-1/2$, but only in a narrow
cone about the preferred axis with opening angle $\sim 1/\gamma$; and
neutrinoless double beta decay is forbidden (and therefore also
unobservable) by lepton number conservation.

Previous authors\cite{Coleman:1998ti,Coleman:1997xq,Colladay:1998fq,%
  Colladay:1996iz} have noted that spurion-mediated Lorentz violation
can lead to two varieties of potentially observable Lorentz-violating
effects: $CPT$ conserving or $CPT$ violating. The same is true for VSR
in its spurion-accessible variants, $T(2)$ and $E(2)$.  However, this
is not necessarily the case for VSR in its $SIM(2)$ avatar.  The $CPT$
operation is equivalent to a \emph{complex} VSR transformation: a
rotation about the $z$-axis by $\pi$ along with an imaginary boost by
the same amount in the $z$-direction.  Thus for amplitudes satisfying
appropriate analyticity properties, $CPT$ follows from $SIM(2)$. (This
argument is similar to the canonical proof of $CPT$ invariance in
Lorentz invariant theories).  While complex $HOM(2)$ can reverse the
sign of any given 4-vector, $CPT$ invariance is not implied because
the necessary transformation is momentum-dependent.

Our paper initiates an exploration of the possibility that the many
empirical successes of special relativity need not demand Lorentz
invariance of the underlying theoretical framework.  Could the Lorentz
invariant Standard Model emerge as an effective theory from a more
fundamental scheme, perhaps operative at the Planck scale, that is VSR
(but not SR) invariant?  Such a scheme, as we have noted, cannot be a
precisely local quantum field theory and its effects, especially for
the case of the spurion-inaccessible variants of VSR, are difficult to
estimate. 

\begin{acknowledgments}
  AGC was supported in part by the Department of Energy under grant
  no. DE-FG02-01ER-40676; SLG by the National Science Foundation under
  grant no. NSF-PHY 0099529.
\end{acknowledgments}

\bibliography{vsr}

\end{document}